# ABOUT TORSIONAL WEYL–DIRAC ELECTRODYNAMICS[1]

## Mark Israelit

Physics Department, University of Haifa-Oranim, Tivon 36006, Israel
israelit@macam.ac.il   israelit@physics.technion.ac.il,

A classical general relativistic theory possessing intrinsic magnetic currents (magnetic monopoles), as well electric ones, and admitting massive photons is built up. As the geometric basis serves a space with torsion and Weylian non-metricity. The theory is coordinate covariant as well as gauge covariant in the Weyl sense. In the limiting case one obtains the ordinary Einstein – Maxwell theory.

## 1. Introduction

Two fundamental electrodynamical phenomena are standing beyond the frames of Maxwell's theory: the magnetic monopole initially introduced [1] by Paul Dirac and the massive photon [2].

Recently the present writer presented a geometrically based classical theory [3] possessing magnetic monopoles and admitting massive photons. In our procedure we made use of the Weyl-Dirac-Rosen framework. The latter originates from Weyl's geometry [4] introduced in order to describe geometrically both, gravitation and electro-magnetism. However, the original Weyl theory had some unsatisfactory features and did not gain general acceptance. In 1973 Dirac modified [5] Weyl's geometry introducing the gauge function $\beta$ and removed earlier difficulties. Later, in the early eighties, the Weyl-Dirac theory was discussed and completed by Nathan Rosen [6]. In the present paper we consider in brief the procedure developed in [3] and discuss some outcomes of the theory.

## 2. The Procedure

To build up the theory we generalize the Weyl-Dirac-Rosen framework. The torsionless Weyl geometry is replaced by a space having both, torsion and Weylian non-metricity. On this broadened basis we have formed a geometrically based action integral, from which the field equations and conservation laws are obtained.

Let us assume that in each point there are given the metric tensor $g_{\mu\nu} = g_{\nu\mu}$, the Weyl connection vector $w_\mu$, the Dirac gauge function $\beta$, as in the original Weyl-Dirac-Rosen framework, and also a torsion tensor $\Gamma^\lambda_{[\mu\nu]}$. Following Dirac [5] we form the action integral of curvature scalars, of Dirac's gauge function $\beta$ and its derivatives. In our case the dynamical variables are: $\Gamma^\lambda_{[\mu\nu]}$, $w_\mu$, $g_{\mu\nu}$, and $\beta$.

Varying in the action the vector $w_\mu$ one obtains the field equation

$$\Phi^{\mu\nu}_{;\nu} = \frac{1}{2}\beta^2(k-6)W^\mu + 2\beta^2\, \Gamma^\alpha_{[\mu\alpha]} + 4\pi J^\mu \qquad (1)$$

---

[1] Contribution to the 8$^{th}$ Marcel Grossmann Meeting on Gen. Rel., Jerusalem, Israel (June 1977). Published in the *MG8 Proceedings* edited by T. Pirani & R. Ruffini. (World Scientific, Singapore, 1999), p. 653.

(an underlined index is to be raised with $g^{\underline{\mu}\nu}$) with $k$ being Dirac's parameter, and a semicolon (;) standing for the covariant partial derivative with respect to the metric $g_{\mu\nu}$. The field strength tensor $\Phi_{\mu\nu}$ in (1) is given by $\Phi_{\mu\nu} = W_{\mu\nu} - 2\Gamma^{\alpha}_{[\mu\nu];\alpha}$, with $W_{\mu\nu}$ standing for the Weylian length curvature tensor $W_{\mu\nu} = w_{\mu;\nu} - w_{\nu;\mu}$. Further $W_{\mu} \equiv w_{\mu} + (\ln\beta)_{,\mu}$ is the gauge-invariant Weyl connection vector, and the electric current density $J^{\mu}$ is introduced as usual: $J^{\mu} \equiv \delta L_m / \delta w_{\mu}$ ($L_m$ is the Lagrangian density of matter).

Considering in the action the variation with respect to $\Gamma^{\lambda}_{[\mu\nu]}$ one has another field equation and combining it with (1) one can express $\Gamma^{\lambda}_{[\mu\nu]}$ in terms of $W_{\mu\nu}$, $J^{\mu}$ and the tensor $\Omega^{[\mu\nu]}_{\lambda}$, given by $16\pi\Omega^{[\mu\nu]}_{\lambda} \equiv \delta L_m / \delta \Gamma^{\lambda}_{[\mu\nu]}$. By this procedure one obtains a dual tensor $\tilde{\Phi}^{\mu\nu}$ satisfying the equation

$$\tilde{\Phi}^{\mu\nu}_{;\nu} = 4\pi L^{\mu} \tag{2}$$

with the intrinsic magnetic current density vector $L^{\mu}$ formed of the $\Omega^{[\mu\nu]}_{\lambda}$-s and with the conservation law $L^{\mu}_{;\mu} = 0$ following from (2).

If one confines oneself to the special purpose of building up the generalized electrodynamics, he can specify the torsion $\Gamma^{\lambda}_{[\mu\nu]}$ taking it as totally antisymmetric

$$\Gamma^{\lambda}_{[\mu\nu]} = -\frac{1}{\sqrt{-g}} g_{\mu\alpha} g_{\nu\beta} \varepsilon^{\lambda\alpha\beta\sigma} V_{\sigma} \tag{3}$$

Where $\varepsilon^{\mu\nu\alpha\beta}$ stands for the completely antisymmetric Levi-Civita symbol with $\varepsilon^{0123} = 1$. If in addition we turn to the Einstein gauge, where $\beta = 1$ and $W_{\mu} \Rightarrow w_{\mu}$, we can write the field strength tensor and its dual explicitly

$$\Phi^{\mu\nu} = \left(w^{\mu}_{;\underline{\nu}} - w^{\nu}_{;\underline{\mu}}\right) - \frac{\varepsilon^{\mu\nu\alpha\sigma}}{\sqrt{-g}}\left(V_{\alpha;\sigma} - V_{\sigma;\alpha}\right)$$

$$\tilde{\Phi}^{\mu\nu} = -2\left(V^{\mu}_{;\underline{\nu}} - V^{\nu}_{;\underline{\mu}}\right) - \frac{\varepsilon^{\mu\nu\alpha\sigma}}{2\sqrt{-g}}\left(w_{\alpha;\sigma} - w_{\sigma;\alpha}\right) \tag{4}$$

and the equations of the electromagnetic field (1) and (2) take the form

$$\Phi^{\mu\nu}_{;\nu} = w^{\mu}_{;\underline{\nu};\nu} - w^{\nu}_{;\underline{\mu};\nu} = -\kappa^2 w^{\mu} + 4\pi J^{\mu}, \text{ and } \tilde{\Phi}^{\mu\nu}_{;\nu} = V^{\mu}_{;\underline{\nu};\nu} - V^{\nu}_{;\underline{\mu};\nu} = -2\pi L^{\mu}. \tag{5}$$

(Dirac's parameter $k$ is replaced by $\kappa^2 = \frac{1}{2}(6-k)$)

The EQ. of the gravitational field in the Einstein gauge may be written as follows

$$G^{\mu\nu} = -8\pi\left(T^{\mu\nu} + \tilde{M}^{\mu\nu} - \overline{M}^{\mu\nu}\right) + 2\kappa^2\left(w^{\mu}w^{\nu} - \frac{1}{2}g^{\mu\nu}w^{\lambda}w_{\lambda}\right) - 2V^{\mu}V^{\nu} - g^{\mu\nu}V^{\lambda}V_{\lambda} \tag{6}$$

where $\tilde{M}^{\mu\nu}$ and $\overline{M}^{\mu\nu}$ are modified energy-momentum density tensors of the electromagnetic field, and $T^{\mu\nu}$ is that of matter. Making use of the Bianchi identity one obtains from (6) the energy-momentum conservation law, and from this follow the EQ-s of motion of charged (either electrically or magnetically) particles.

For a moment let us go back to (5) and consider a current-free small region, in which the curvature is negligible. Then we obtain

$$w^{\mu}_{;\nu;\nu} + \kappa^2 w^{\mu} = 0 \tag{7}$$

From the quantum mechanical point of view (7) describes a particle with spin 1, and mass given in conventional units by

$$m_{\gamma} = \left(\frac{\hbar}{c}\right)\kappa \equiv \left(\frac{\hbar}{c}\right)\sqrt{\frac{6-k}{2}} \tag{8}$$

Thus the field is represented by massive photons.

## 3. Discussion

In the presented torsional Weyl-Dirac electrodynamics the strength tensor $\Phi_{\mu\nu}$ of the electromagnetic field is formed of two parts, the Maxwellian one, and the divergence of the totally antisymmetric torsion tensor. The dual field tensor $\tilde{\Phi}_{\mu\nu}$ has a non-vanishing divergence, so that according to the field EQ-s there exists an intrinsic magnetic current density vector in addition to the electric one. The Dirac parameter $k$ was not fixed by $k = 6$, and as a result Proca-like terms appear in the EQ. of the electromagnetic field (5), so that one has massive field quanta, photons (7), (8).

In this theory an intrinsic magnetic field generates massive photons, while in the absence of magnetic fields the photon is massless. We have investigated the equations of motion in the Einstein gauge. It turns out that on electrically charged test particles act electric, as well as magnetic, currents. Further the electric-electric and magnetic-electric interactions may be transmitted either by massive, or by massless photons. However, considering a magnetically charged test particle we found that the magnetic-magnetic interaction is transmitted by massive photons. Thus, on the one hand the photon is massive only in the presence of magnetic fields, on the other hand the massive photon is responsible for the interaction between magnetic monopoles.

A static spherically symmetric solution was obtained. From this one obtains either the metric, and the field of a magnetic monopole, ore that of an electric one, the Reissner-Nordstrøm metric. Hence, a magnetic monopole cannot be located together with an electric one. It is also shown that the magnetic monopole is massive, and that it is surrounded by a spherical surface on which the metric is singular.

Details may be found in [3, 7].